\documentclass[journal]{IEEEtran}

\usepackage{mathrsfs}

\usepackage{cite}
\usepackage{graphicx}
\ifCLASSINFOpdf
\else
\fi
\usepackage{todonotes}
\usepackage{amsmath,mathtools}
\usepackage{amssymb}
\usepackage{amsfonts}

\ifCLASSOPTIONcompsoc
 \usepackage[caption=false,font=normalsize,labelfont=sf,textfont=sf]{subfig}
\else
  \usepackage[caption=false,font=footnotesize]{subfig}
\fi

\usepackage{xcolor}
\begin{document}
%
\title{Asymptotic MIMO Channel Model for Diffusive MC with Fully-absorbing Receivers}
%
%
%

\author{\IEEEauthorblockN{Fardad~Vakilipoor\IEEEauthorrefmark{3}~\IEEEmembership{Student Member,~IEEE,}
Marco~Ferrari\IEEEauthorrefmark{1}~\IEEEmembership{Member,~IEEE,}
Maurizio~Magarini\IEEEauthorrefmark{3}~\IEEEmembership{Member,~IEEE,}}
\IEEEauthorrefmark{3}Dipartimento di Elettronica, Informazione e Bioingegneria, Politecnico di Milano, 20133 Milano, Italy\\
\IEEEauthorblockA{\IEEEauthorrefmark{1}CNR - IEIIT, DEIB, Politecnico di Milano, 20133 Milano, Italy\\
Email: fardad.vakilipoor@polimi.it,
marco.ferrari@ieiit.cnr.it,
maurizio.magarini@polimi.it
}
}
\maketitle
\begin{abstract}
This letter introduces an analytical model that gives the asymptotic cumulative number of molecules absorbed by spherical receivers in a diffusive multiple-input multiple-output (MIMO) molecular communication (MC) system with pointwise transmitters. In the proposed model, the reciprocal effect among the fully absorbing (FA) receivers is described by using the concept of fictitious pointwise negative source of molecules, the best position of which for each spherical receiver being defined by its \textit{absorption barycenter}. 
We show that there is an agreement between the proposed asymptotic model and the numerical solution of the exact analytical model from the literature that describes the interaction among the receivers, which is solved for a sufficient long time. We resort to numerical solution because for the exact model there is no analytical solution apart for the case of one transmitter and two receivers. We demonstrate that the barycenter tends to coincide with the center of the FA receivers in the temporal asymptotic case.
\end{abstract}
\vspace{-.2cm}
%
\IEEEpeerreviewmaketitle
\vspace{-.3cm}
\section{Introduction}
%
%
%
\IEEEPARstart{M}{olecular} communication (MC) is a bio-inspired communication paradigm where signals are exchanged between nanoscale devices by using particles, i.e. molecules, as carriers of information~\cite{vakilipoor2021low}. Applications of MC systems mainly revolve around the health sector, such as targeted drug delivery, nanomedicine, etc.~\cite{chude2017molecular,akyildiz2019moving,zhao2021release}. Nanomachines are the most integral units in MC systems that are only capable of performing primary functions like sensing, actuation, and computation. They are unable to do complicated tasks and, for this reason, they need to establish reliable cooperation in nanonetworks to perform complex tasks. 
\par In nanonetworks there are multiple transmitters and multiple receivers involved in the communication process~\cite{bi2021survey}. Although nanonetworks are characterized by the presence of multiple transmitting and receiving nanomachines~\cite{lee2017machine}, so far research in MC has mainly focused on systems with a single transmitter and a single receiver. This highlights the importance of studying and defining new models for multiple-input multiple-output (MIMO) systems in MC. 

Receivers in MC can be basically divided into two different categories~\cite{farsad2016comprehensive,jamali2019channel}. The first is defined by passive receivers, where there is no interaction with information molecules. The second by active receivers, in which there is reciprocal interaction that can occur through an absorption process or via chemical reactions. Thus, the case of multiple active receivers requires extra consideration due to their interaction and, for this reason, in this letter we focus on a diffusive MIMO MC scenario with fully absorbing (FA) spherical receivers. 

\par The diffusive MIMO MC system was already considered in the literature~\cite{koo2016molecular,liu2021modeling,kwak2020two}.
Recently, in~\cite{Fardad1} an analytical model has been introduced to describe the channel impulse response between a pointwise transmitter and a given number of FA receivers. The approach is based on the idea of describing the effect of all the FA receivers, except the one considered for the derivation of the impulse response, as pointwise sources of negative molecules.
A key contribution of~\cite{Fardad1} is to show that the position of the pointwise source of negative molecules is not necessarily the center of the receivers and that its position varies according to the time of observation. The \textit{absorption barycenter}
, depending on the relative position of receivers and transmitter, is proposed to locate the negative source.
\par By following the analysis done in~\cite{Fardad1}, we model the diffusive MIMO MC scenario by means of a system of equations and derive an analytical expression for the asymptotic value of the cumulative number of molecules absorbed by each FA receiver. 
To our best knowledge, this is the first work that gives an analytical expression to describe the asymptotic behavior of a diffusive MIMO MC system with FA receivers. 
\par The letter is organized as follows. Section~\ref{sec:system_model} introduces the baseline diffusive MC scenario with a single pointwise transmitter and a variable number of FA spherical receivers.
In Sec.~\ref{sec:Asymptotic_analysis} we derive the asymptotic behavior of the reference scenario given in Sec.~\ref{sec:system_model}, in terms of asymptotic cumulative number of molecules absorbed by each receiver, and then extend it to the MIMO case. Section~\ref{sec:simulation_results} presents the results and validates the proposed asymptotic model through a comparison with the numerical solution of the exact analytical model given in the literature. Concluding remarks are provided in Sec.~\ref{sec:conclusion}.

\section{System model}\label{sec:system_model}
\begin{figure}[!t]
    \centering
    \includegraphics[width=0.85\columnwidth]{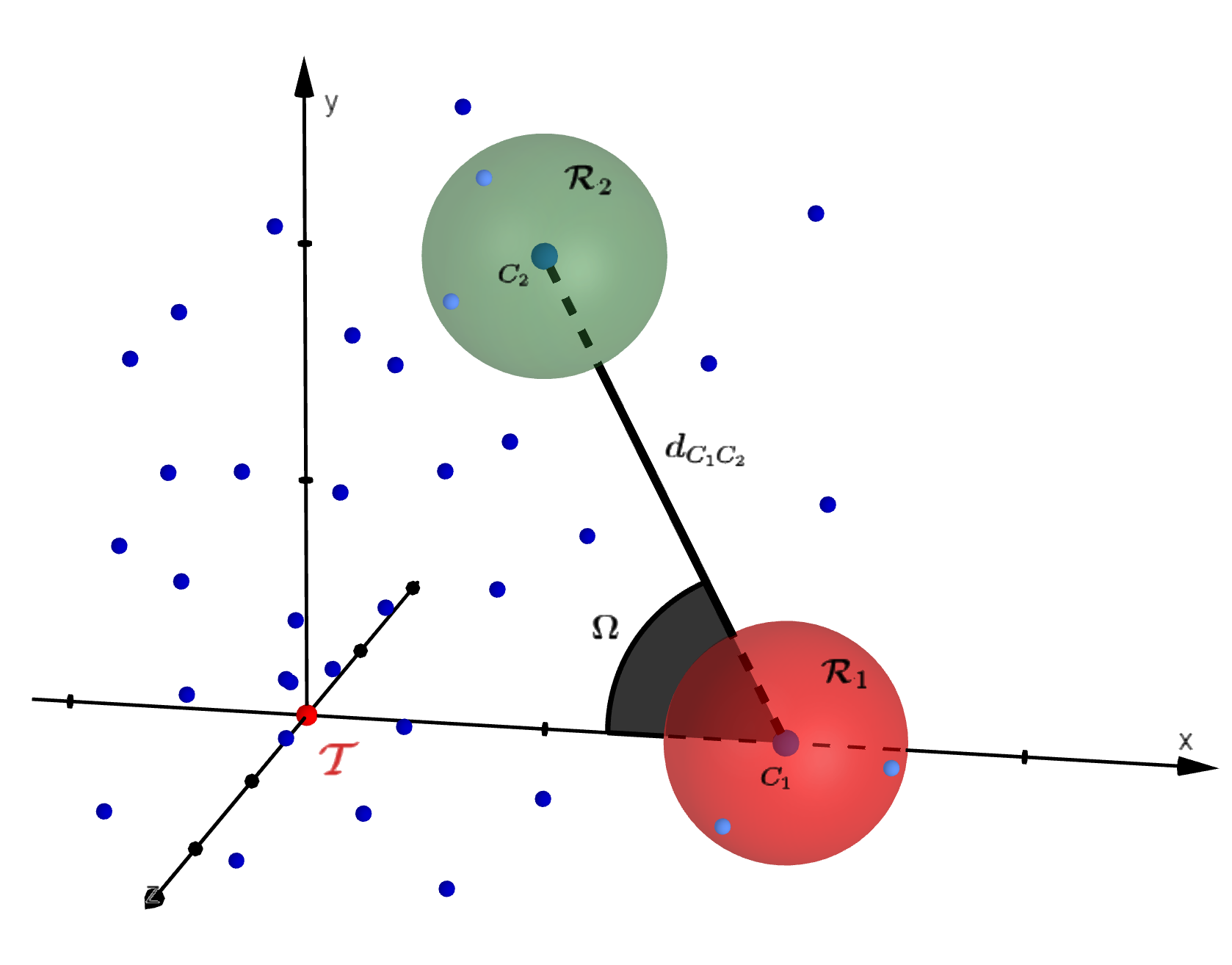}
    \caption{MCvD system with two FA receivers centered at points $C_1$ and $C_2$.}
    \label{fig:topology}
\end{figure}

The system considered in this letter consists of pointwise transmitters, spherical FA receivers, and diffusive channel. The transmitters are dimensionless and each of them emits $N_{T}$ messenger molecules of the same type in the environment instantaneously. Molecules diffuse through the medium with a constant diffusion coefficient $D$ in an unbounded three-dimensional (3D) environment. Receivers absorb messenger molecules that hit their surface and trap them. As a reference, in Fig.~\ref{fig:topology} the example of one transmitter and two receivers is shown. The FA characteristic introduces a coupling effect among the receivers' observations that can be interpreted as a reduction of molecules from the environment~\cite{Fardad1}. Thus, we need to consider the reciprocal impact to study the number of molecules absorbed by each receiver. 
In what follows, we first review the single-input single-output (SISO) case, then we develop the single-input two-output (SITO) scenario, and finally generalize it to single-input multiple-output (SIMO).

\subsection{Single Input Single Output (SISO)}

Diffusive propagation of molecules is governed by Fick's second law that links the time derivative of the flux to the Laplacian of the molecules concentration $p\left(r,t\right)$ at distance $r$ and time $t$ as~\cite{farsad2016comprehensive}
\begin{equation}
    \frac{\partial p \left( r,t \right)}{\partial t} = D \nabla^2p
    \left( r,t \right). \label{eq:2nd Fick}
\end{equation}
The initial and boundary conditions of~\eqref{eq:2nd Fick} vary subject to the MC system characterization. Yilmaz~\textit{et al.}~\cite{yilmaz2014three} specified the boundary and initial conditions for an impulsive release of molecules and an FA receiver $\mathcal{R}_{1}$. They obtained an analytical expression that describes the hitting rate of the molecules onto the receiving cell surface, namely $f\left(d_{1},t\right)$, which depends on the distance~$d_{1}$ between the transmitter $\mathcal{T}$ and the center of the receiver $\mathcal{R}_{1}$, at time~$t$. 
The channel impulse response of a diffusive MC channel with a single spherical FA receiver of radius $R$, centered at distance $d_{1}$ from transmitter reads
\begin{equation}
f \left( d_1,t \right) = \frac{R \left(d_{1}-R\right)}{d_{1}\sqrt{4 \pi D
t^3}} e^{-\frac{\left(d_{1}-R \right)^2}{4Dt}}. 
\label{eq:imp_SISO}
\end{equation}
The absorption rate,~\textit{i.e.}, the number of molecules absorbed by the cell per unit time, is given by
\begin{equation}
n_{1} \left( t \right) = N_T f \left( d_{1},t \right).
\label{eq:n_R(t)}
\end{equation}
The overall number of absorbed molecules is obtained from integration of~\eqref{eq:n_R(t)} up to time $t$
     \begin{align}
    N_{1}(t) =  \int_{0}^t n_{1} \left( u \right) \,du 
    =  \frac{N_{T}R}{d_1} \mathrm{erfc}\left ( \frac{d_1-R}{2\sqrt{Dt}} \right ),
    \label{eq:integ}\end{align} 
where 
\begin{equation}
    \mathrm{erfc}\left ( z \right ) = 1 - \frac{2}{\sqrt{\pi}}\int_{0}^{z} e^{-\tau ^2}d\tau ,
\end{equation}
is the complementary error function.
\subsection{Single Input Two Output (SITO)}\label{sec_SITO}
As shown in~\cite{Fardad1}, the reciprocal effect among FA receivers can be obtained by introducing the concept of negative pointwise source of molecules, to account for the amount of molecules absorbed by each receiver and hence removed from the surrounding space. Its position is well described by the barycenter that can be associated with each of the FA receivers. The barycenter of each receiver is defined as the spatial average of its surface points hit by molecules, given geometry of the problem.
We can write the absorption rate of each receiver as~\cite[eq. (20)]{Fardad1}
\begin{equation}
    \begin{cases}
        n_1\left(t\right) = N_T f_{1}  - n_2\left(t\right) \star f_{1,2} , 
        \\
        n_2\left(t\right) = N_T f_{2}  - n_1\left(t\right) \star f_{2,1} ,
    \end{cases}
    \label{eq:B_SE}
\end{equation}
where $\star$ is the convolution, $f_{1}$$\,=\,$$f\left( d_1,t \right)$, $f_{1,2}$$\,=\,$$f\left( d_{1,2},t \right)$, and $d_{1,2}$ is the distance between the center of $\mathcal{R}_1$ and barycenter of $\mathcal{R}_2$. On the other hand, $f_{2}$$\,=\,$$f\left( d_2,t \right)$, $f_{2,1}$$\,=\,$$f\left( d_{2,1},t \right)$, and $d_{2,1}$ is the distance between the center of $\mathcal{R}_2$ and barycenter of $\mathcal{R}_1$. In order to find the number of molecules absorbed by each receiver, it is required to integrate and solve~\eqref{eq:B_SE}. We take the Laplace transform of the integral of~\eqref{eq:B_SE}
 \begin{equation}
    \begin{cases}
        \hat{N}_1\left(s\right) = \frac{N_T \hat{f}_1}{s}   - \hat{N}_2\left(s\right)  \hat{f}_{1,2} ,
        \\
        \hat{N}_2\left(s\right) = \frac{N_T \hat{f}_{2}}{s}   - \hat{N}_1\left(s\right)  \hat{f}_{2,1} , 
    \end{cases}
    \label{eq:SITO_S}
\end{equation}
where $\mathscr{L}\{f\}$$\,=\,$$\hat{f}$ and $\mathscr{L}\{N\}$$\,=\,$$\hat{N}$. After rearranging~\eqref{eq:SITO_S} we can write it as matrix multiplication
\begin{equation}
    \begin{bmatrix}
        \frac{N_T\hat{f}_{1}}{s} \\ 
        \frac{N_T\hat{f}_{2}}{s}
    \end{bmatrix}
    =
    \begin{bmatrix}
        1 & \hat{f}_{1,2}\\ 
        \hat{f}_{2,1} & 1 
    \end{bmatrix}
    \begin{bmatrix}
        \hat{N}_{1}\left(s \right )\\ 
        \hat{N}_{2}\left(s \right )
    \end{bmatrix}.\label{eq:1*2Mat}
\end{equation}
Thus, the solution in $s$ domain is obtained by a matrix inversion followed by multiplication
\begin{equation}
    \begin{bmatrix}
        \hat{N}_{1}\left(s \right )\\ 
        \hat{N}_{2}\left(s \right )
    \end{bmatrix}
    =
    \begin{bmatrix}
        1 & \hat{f}_{1,2}\\ 
        \hat{f}_{2,1} & 1 
    \end{bmatrix}^{-1}
    \begin{bmatrix}
        \frac{N_T\hat{f}_{1}}{s} \\ 
        \frac{N_T\hat{f}_{2}}{s}
    \end{bmatrix}.
    \label{eq:1*2Mat_Inv}
\end{equation}
By computing the inverse Laplace transform of~\eqref{eq:1*2Mat_Inv} a closed-form solution can be obtained that expresses the cumulative number of absorbed molecules by $\mathcal{R}_{1}$ as~\cite[eq. (21)]{Fardad1} 
\begin{align}
     \resizebox{.89\hsize}{!}{$\begin{aligned}
    N_1(t) &= \frac{N_{T}R}{d_1}\sum_{n=0}^{\infty}\frac{R^{2n}}{(d_{1,2}d_{2,1})^{n}} \mathrm{erfc}\left(\frac{(d_1-R)+n(d_{1,2}+d_{2,1}-2R)}{2\sqrt{Dt}}\right)\\
    &\hspace{0.4cm}- \frac{N_{T}R^{2}}{d_{1,2}d_2}\sum_{n=0}^{\infty}\frac{R^{2n}}{(d_{1,2}d_{2,1})^{n}} \mathrm{erfc}\left(\frac{(d_{1,2}+d_2-2R)+n(d_{1,2}+d_{2,1}-2R}{2\sqrt{Dt}}\right) .
     \end{aligned}$}\label{eq:SITO_N1_detD}
 \end{align}
\subsection{Single-Input Multiple-Output (SIMO)}
By following the same strategy seen in Sec.~\ref{sec_SITO} for the SITO case, one can write the generalized form of~\eqref{eq:SITO_S} as
 \begin{equation}\label{eq:SIMO_S}
 \resizebox{0.88\hsize}{!}{$
    \begin{cases}
        \hat{N}_1\left(s\right) = \frac{N_T \hat{f}_1}{s}   - \hat{N}_2\left(s\right)  \hat{f}_{1,2}  -\hat{N}_3\left(s\right)  \hat{f}_{1,3} \cdots - \hat{N}_p\left(s\right)  \hat{f}_{1,p}
        \\
        \hat{N}_2\left(s\right) = \frac{N_T \hat{f}_{2}}{s}   - \hat{N}_1\left(s\right)  \hat{f}_{2,1} - \hat{N}_3\left(s\right)  \hat{f}_{2,3} \cdots - \hat{N}_p\left(s\right)  \hat{f}_{2,p}
        \\
        \vdots
        \\
        \hat{N}_p\left(s\right) = \frac{N_T \hat{f}_{p}}{s}   - \hat{N}_1\left(s\right)  \hat{f}_{p,1} - \hat{N}_2\left(s\right)  \hat{f}_{p,2} \cdots - \hat{N}_{p-1}\left(s\right)  \hat{f}_{p,p-1}
    \end{cases},%
 $   }
\end{equation}
Mimicking the same procedure as before, the solution of~\eqref{eq:SIMO_S} in $s$ domain can be written as
\begin{equation}
\resizebox{0.88\hsize}{!}{$
    \begin{bmatrix}
        \hat{N}_{1}\left(s \right )\\ 
        \hat{N}_{2}\left(s \right )\\
        \vdots\\
        \hat{N}_{p}\left(s \right )
    \end{bmatrix}
    =
    \begin{bmatrix}
        1 & \hat{f}_{1,2} & \hat{f}_{1,3} & \hdots & \hat{f}_{1,p}\\ 
        \hat{f}_{2,1} & 1 & \hat{f}_{2,3} & \hdots & \hat{f}_{2,p}\\
        \hat{f}_{3,1} & \hat{f}_{3,2} & 1 & \hdots & \hat{f}_{3,p}\\
        \vdots & \vdots & \vdots & \ddots & \vdots\\
        \hat{f}_{p,1} & \hat{f}_{p,2} & \hat{f}_{p,3} &\hdots & 1
    \end{bmatrix}^{-1}
    \begin{bmatrix}
        \frac{N_T\hat{f}_{1}}{s} \\ 
        \frac{N_T\hat{f}_{2}}{s} \\
        \vdots\\
        \frac{N_T\hat{f}_{p}}{s}
    \end{bmatrix} .
    \label{eq:1*pMat}
    $}
\end{equation}
Unlike the SITO case, the time domain closed-form solution of~\eqref{eq:1*pMat} has not been derived yet. 
Note that, the system of equations~\eqref{eq:SIMO_S} was originally a time domain integration, before applying the Laplace transform, and it can be solved numerically as shown in~\cite{Fardad1}.
\begin{figure*}[t!]
    \centering
  \subfloat[\label{fig:a}]{%
       \includegraphics[width=0.33\linewidth]{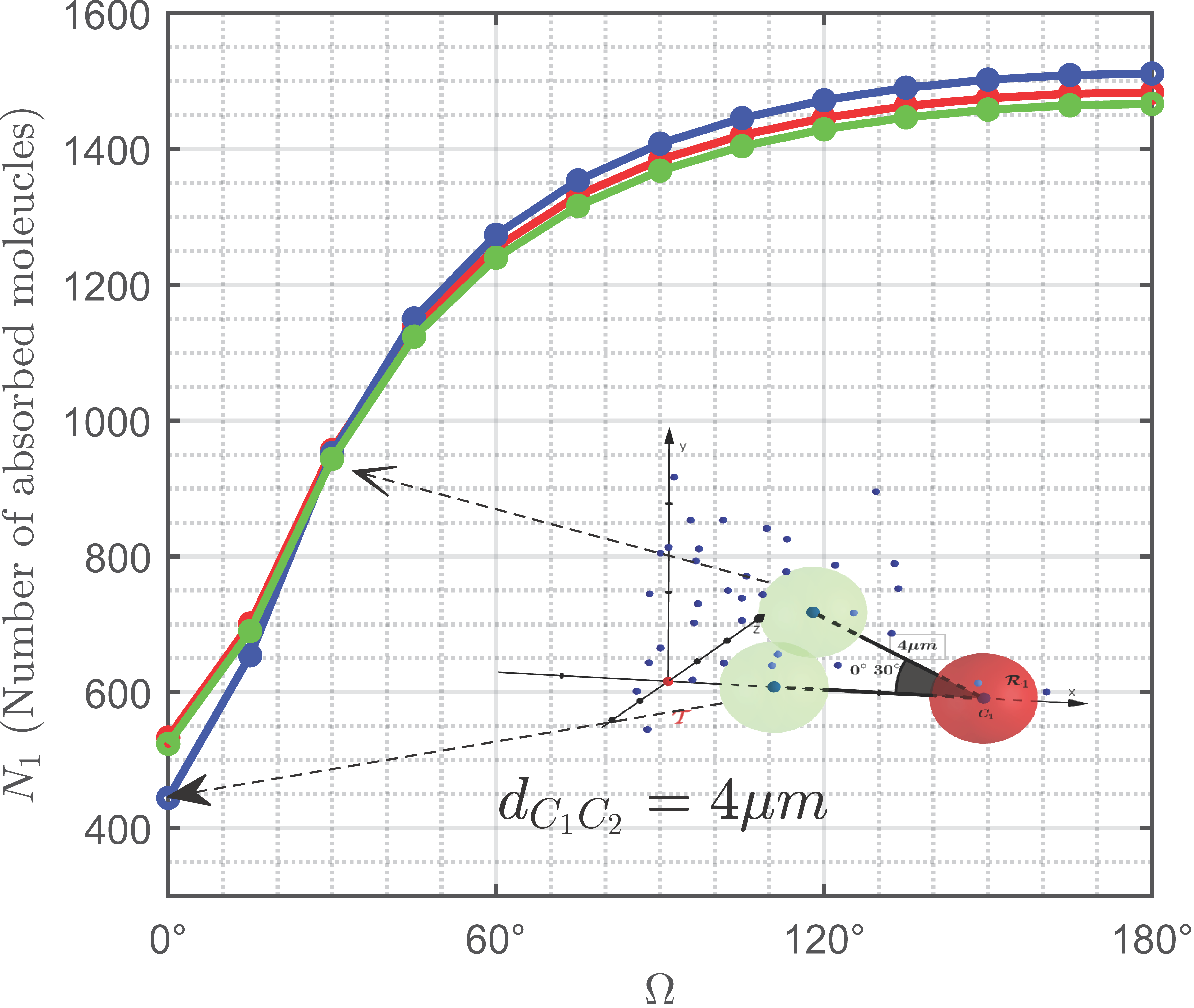}}
  \subfloat[\label{fig:b}]{%
        \includegraphics[width=0.31\linewidth]{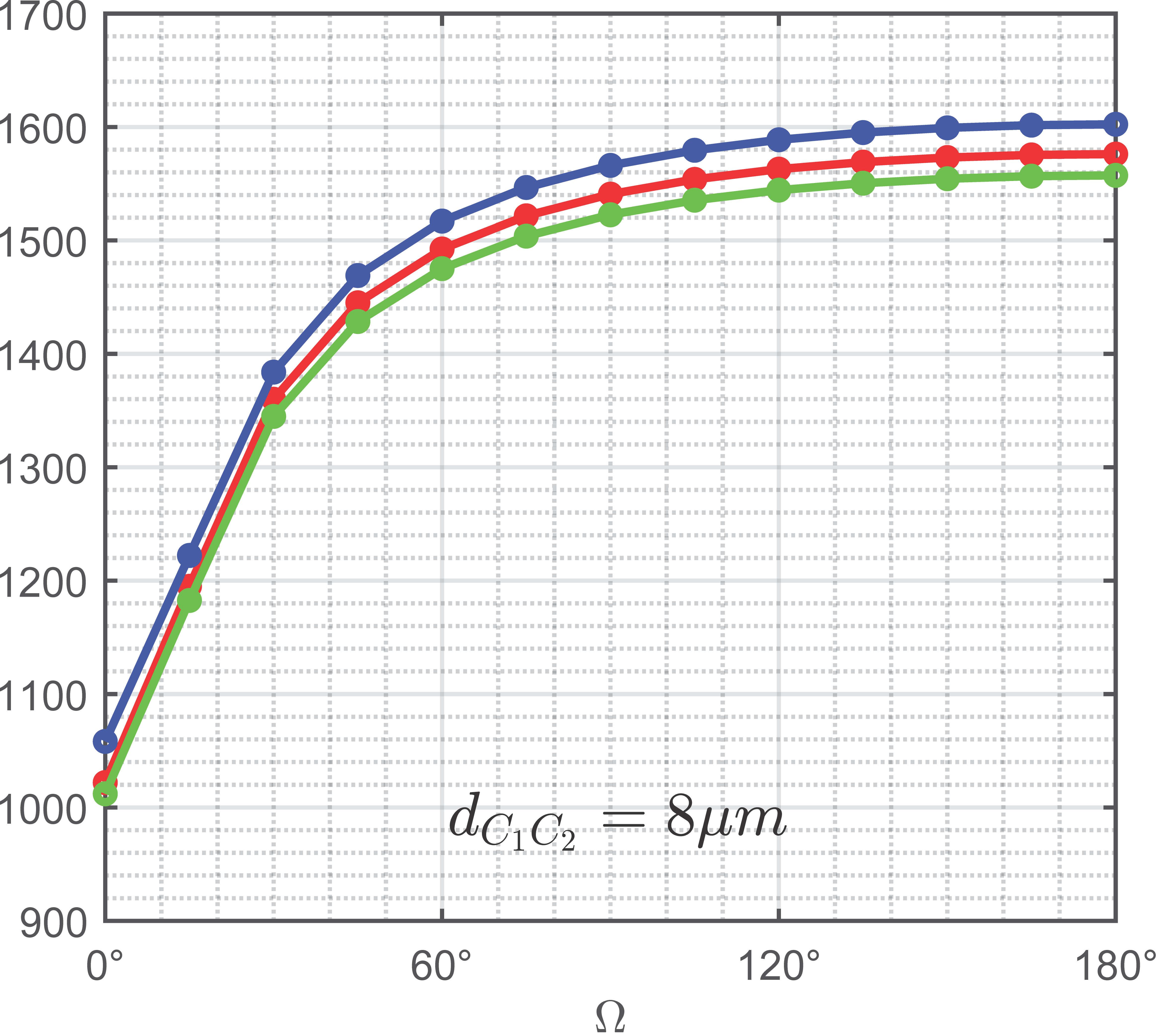}}
  \subfloat[\label{fig:c}]{%
        \includegraphics[width=0.31\linewidth]{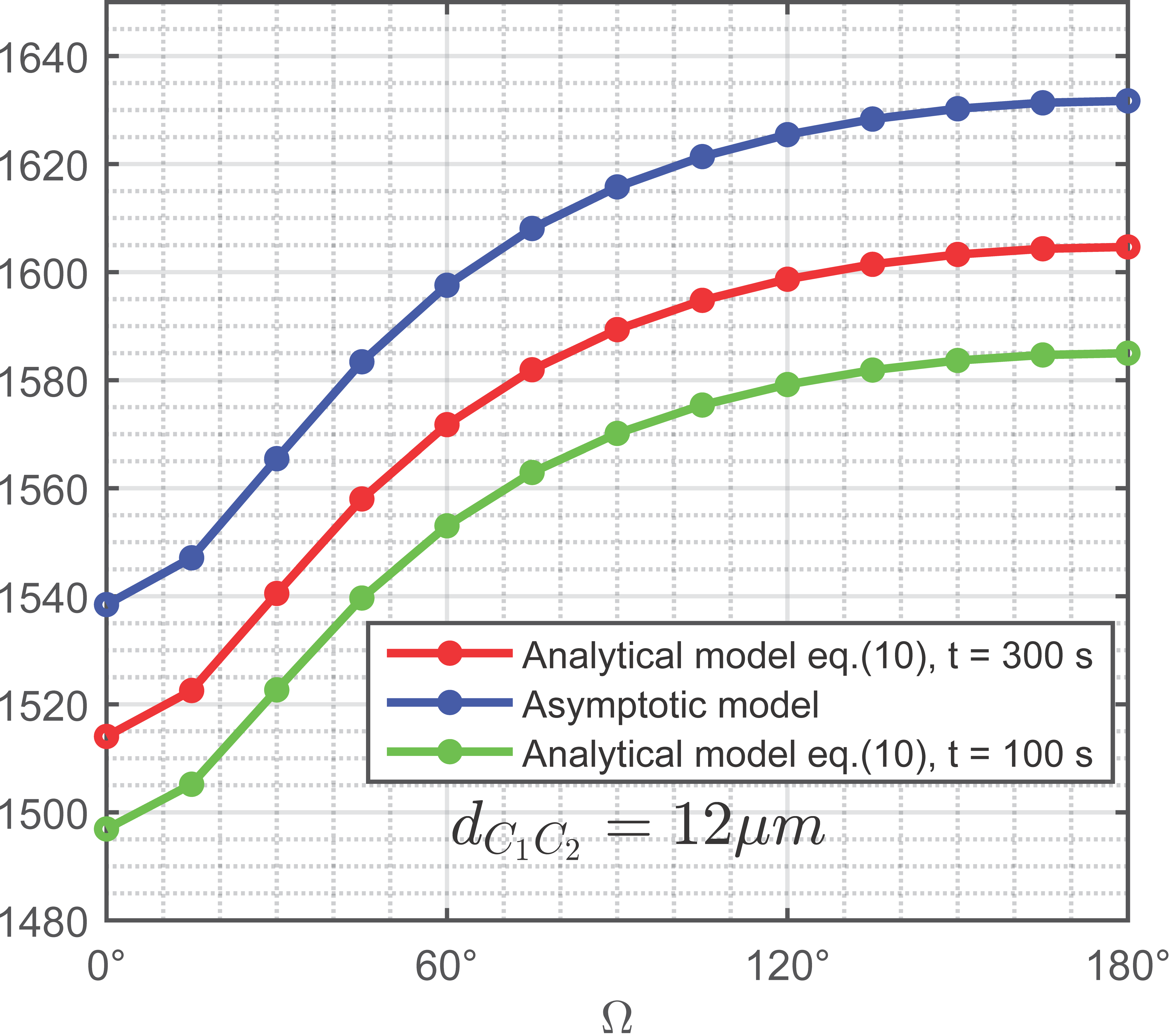}}
  \caption{Cumulative expected number of molecules $N_1(t)$ absorbed by $\mathcal{R}_1$ after $t$$\,\in\,$$\{100,300\}$ seconds, in the scenario of Fig.~\ref{fig:topology} with $d_1$$\,=\,$$6\,\mu\mathrm{m}$, for various positions of $\mathcal{R}_2$ identified by $\Omega$ and $d_{C_1C_2}=\{4,8,12\}\,\mu$m in (a), (b), and (c), respectively.}
  \label{fig:Sim_result} 
\end{figure*}
\section{Asymptotic analysis}\label{sec:Asymptotic_analysis}
The main aim of this letter is to explore the asymptotic behavior,~$t\to \infty$, of FA receivers observation. The asymptotic analysis is mainly used to study the interaction among receivers. Note that, since we are investigating the asymptotic behavior of the model, the barycenters can be considered located at the centers of the spherical FA receivers, as shown in~\cite{Fardad1}. Hence, $d_{i,j}$ represents the distance between the centers of $\mathcal{R}_i$ and $\mathcal{R}_j$. In the following, we analyze the asymptotic behavior of the three scenarios discussed in the previous section and, finally, we extend the SIMO model to MIMO.
\subsection{SISO}
The asymptotic number of molecules absorbed by $\mathcal{R}_1$ in the SISO case is calculated by taking the limit of~\eqref{eq:integ} as
\begin{align}
    \lim_{t \to \infty} N_{1}\left(t\right)  = \lim_{t \to \infty} \frac{N_{T}R}{d_1} \mathrm{erfc}\left ( \frac{d_1-R}{2\sqrt{Dt}} \right ) =\frac{N_{T}R}{d_{1}} .\label{eq:SISO_Asm}
\end{align} 
\subsection{SITO}
The asymptotic value in this case is obtained by rewriting~\eqref{eq:SITO_N1_detD} as 
 \begin{align}
     \resizebox{.89\hsize}{!}{$\begin{aligned}
     \lim_{t \to \infty} N_{1}\left(t\right) & = \sum_{n=0}^{\infty}  \Bigg(\frac{R^{2}}{d_{C_{R}S_{I}1,2}d_{2,1}}\Bigg)^{n} \lim_{t \to \infty} \Bigg[ \frac{N_{T}R}{d_1} \mathrm{erfc}\left(\frac{(d_1-R)+n(d_{1,2}+d_{2,1}-2R)}{2\sqrt{Dt}}\right) \\
    &\hspace{0.4cm}- \frac{N_{T}R^{2}}{d_{1,2}d_2}  \mathrm{erfc}\left(\frac{(d_{1,2}+d_2-2R)+n(d_{1,2}+d_{2,1}-2R}{2\sqrt{Dt}}\right)\Bigg] .
     \end{aligned}$}\label{eq:SITO_limit_T_singleSUM}
 \end{align}
Since the value of $\mathrm{erfc}$ converges to one as the argument goes to zero,~\eqref{eq:SITO_limit_T_singleSUM} simplifies to
\begin{equation}
     \lim_{t \to \infty} N_{1}\left(t\right)  = \sum_{n=0}^{\infty}  \Bigg(\frac{R^{2}}{d_{1,2}d_{2,1}}\Bigg)^{n}  \left(\frac{N_{T}R}{d_1} - \frac{N_{T}R^{2}}{d_{1,2}d_2}\right).
\end{equation}
Assuming $\frac{R^{2}}{d_{1,2}d_{2,1}}$$\,<\,$$1$,~\textit{i.e.}, receivers do not touch nor interfere each other, by using limit of the geometrical series we get
 \begin{equation}
 \resizebox{0.88\hsize}{!}{$
      \lim_{t \to \infty} N_{1}\left(t\right)  = N_{T}R\left(\frac{d_{1,2}d_2 - Rd_{1}}{d_{1}d_{2}d_{1,2}}\right) \left(\frac{d_{1,2}d_{2,1}}{d_{1,2}d_{2,1} - R^{2}}\right) .\label{eq:SITO_ASYM}
      $}
 \end{equation}
It is expected that in a SITO scenario if $\mathcal{R}_2$ is located far from $\mathcal{R}_1$, then $N_1$ must behave as there was no other FA receivers around. This characteristic is perfectly observable in our derivations. In fact, if we fix the position of $\mathcal{R}_1$ and move $\mathcal{R}_2$ far away, \textit{i.e.},~$d_{2},d_{1,2},d_{2,1}\to \infty$,~\eqref{eq:SITO_ASYM} converges to~\eqref{eq:SISO_Asm}. It is worth noting that we can obtain the same expression as~\eqref{eq:SITO_ASYM} by using the finite value theorem and applying it to~\eqref{eq:1*2Mat_Inv}.
\subsection{SIMO}
For the SIMO scenario we hire~\eqref{eq:1*pMat} and take advantage of finite value theorem, which reads
\begin{equation}\label{eq:FVT}
    \lim_{t \to \infty} \begin{bmatrix}
        {N}_{1}\left(t \right )\\ 
        \vdots\\ 
        {N}_{p}\left(t \right )
    \end{bmatrix}
    = \lim_{s \to 0}s \begin{bmatrix}
        \hat{N}_{1}\left(s \right )\\ 
        \vdots\\ 
        \hat{N}_{p}\left(s \right )
    \end{bmatrix}.
\end{equation}
Substituting the right-hand side of~\eqref{eq:1*pMat} in~\eqref{eq:FVT} we get
\begin{equation} \label{eq:SIMO_Lim}
\resizebox{0.88\hsize}{!}{$
     \lim_{t \to \infty}\resizebox{.16\hsize}{!}{$ \begin{bmatrix}
        {N}_{1}\left(t \right )\\ 
        {N}_{2}\left(t \right )\\
        \vdots\\
        {N}_{p}\left(t \right )
    \end{bmatrix}
    =$}
    \lim_{s \to 0}\resizebox{.67\hsize}{!}{$ \begin{bmatrix}
        1 & \hat{f}_{1,2} & \hat{f}_{1,3} & \hdots & \hat{f}_{1,p}\\ 
        \hat{f}_{2,1} & 1 & \hat{f}_{2,3} & \hdots & \hat{f}_{2,p}\\
        \hat{f}_{3,1} & \hat{f}_{3,2} & 1 & \hdots & \hat{f}_{3,p}\\
        \vdots & \vdots & \vdots & \ddots & \vdots\\
        \hat{f}_{p,1} & \hat{f}_{p,2} & \hat{f}_{p,3} &\hdots & 1
    \end{bmatrix}^{\mathrlap{-1}}
        \begin{bmatrix}
        N_T\hat{f}_{1} \\
        N_T\hat{f}_{2} \\
        \vdots \\
        N_T\hat{f}_{p}
    \end{bmatrix}$}.
    $}
 \end{equation}
As a last step we need to find the limit of $\hat{f}$ as $s\to 0$. By taking the Laplace transform of $f$, according to~\cite{schiff1999laplace}, and by computing the limit we get
\begin{align}
     \lim_{s \to 0}\hat{f}(d,s) = \lim_{s \to 0} \frac{R}{d}e^{-\frac{d-R}{\sqrt{D}}\sqrt{s}}
     = \frac{R}{d}.
     \label{eq:f_hat_lim}
 \end{align}
We can substitute the result from~\eqref{eq:f_hat_lim} inside~\eqref{eq:SIMO_Lim}
 \begin{equation}
 \resizebox{0.88\hsize}{!}{$
     \lim_{t \to \infty} \begin{bmatrix}
        {N}_{1}\left(t \right )\\ 
        {N}_{2}\left(t \right )\\
        \vdots\\
        {N}_{p}\left(t \right )
    \end{bmatrix}
    =
    \underbrace{
    \begin{bmatrix}
        1 & \frac{R}{d_{1,2}} & \frac{R}{d_{1,3}} & \hdots & \frac{R}{d_{1,p}}\\ 
        \frac{R}{d_{2,1}} & 1 & \frac{R}{d_{2,3}} & \hdots & \frac{R}{d_{2,p}}\\
        \frac{R}{d_{3,1}} & \frac{R}{d_{3,2}} & 1 & \hdots & \frac{R}{d_{3,p}}\\
        \vdots & \vdots & \vdots & \ddots & \vdots\\
        \frac{R}{d_{p,1}} & \frac{R}{d_{p,2}} & \frac{R}{d_{p,3}} &\hdots & 1
    \end{bmatrix}^{\mathrlap{-1}}}_{\text{$\pmb{\mathscr{R}}$}}
    \underbrace{\begin{bmatrix}
        N_T\frac{R}{d_{1}} \\
        N_T\frac{R}{d_{2}} \\
        \vdots \\
        N_T\frac{R}{d_{p}}
    \end{bmatrix}}_{\text{$\pmb{\mathscr{T}}$}}.
$}
 \end{equation}
Matrices~$\pmb{\mathscr{R}}$ and $\pmb{\mathscr{T}}$ are dependent on the distance between the receivers themselves and the distance between receivers and transmitter, respectively.
\subsection{Extension to MIMO}
All transmitters are assumed to be pointwise. Considering their independency, we can generalize matrix~$\pmb{\mathscr{T}}$ to multiple transmitters by introducing a new notation for the distance between each receiver and transmitter. We define the distance between~$\mathcal{T}_{i}$ and~$\mathcal{R}_{j}$ by $\prescript{}{i}{d}_{j}$. Imposing the superposition effect on~$\pmb{\mathscr{T}}$, we can extend it to the case of $q$ transmitters as
\begin{equation}
    \pmb{\mathscr{T}} = 
    \begin{bmatrix}
        N_T\sum_{i=1}^{i=q}\frac{R}{\prescript{}{i}{d}_{1}} \\
        N_T\sum_{i=1}^{i=q}\frac{R}{\prescript{}{i}{d}_{2}} \\
        \vdots \\
        N_T\sum_{i=1}^{i=q}\frac{R}{\prescript{}{i}{d}_{p}}
    \end{bmatrix}.
\end{equation}
The asymptotic value of the number of absorbed molecules in a 3D diffusive molecular communication system consisting of $p$ spherical FA receivers and $q$ pointwise transmitters is 
 \begin{equation} \label{eq:MIMO_Asm}
 \resizebox{0.88\hsize}{!}{$
     \lim_{t \to \infty} \begin{bmatrix}
        {N}_{1}\left(t \right )\\ 
        {N}_{2}\left(t \right )\\
        \vdots\\
        {N}_{p}\left(t \right )
    \end{bmatrix}
    =
    \resizebox{.67\hsize}{!}{$\begin{bmatrix}
        1 & \frac{R}{d_{1,2}} & \frac{R}{d_{1,3}} & \hdots & \frac{R}{d_{1,p}}\\ 
        \frac{R}{d_{2,1}} & 1 & \frac{R}{d_{2,3}} & \hdots & \frac{R}{d_{2,p}}\\
        \frac{R}{d_{3,1}} & \frac{R}{d_{3,2}} & 1 & \hdots & \frac{R}{d_{3,p}}\\
        \vdots & \vdots & \vdots & \ddots & \vdots\\
        \frac{R}{d_{p,1}} & \frac{R}{d_{p,2}} & \frac{R}{d_{p,3}} &\hdots & 1
    \end{bmatrix}^{\mathrlap{-1}}
    \begin{bmatrix}
        N_T\sum_{i=1}^{i=q}\frac{R}{\prescript{}{i}{d}_{1}} \\
        N_T\sum_{i=1}^{i=q}\frac{R}{\prescript{}{i}{d}_{2}} \\
        \vdots \\
        N_T\sum_{i=1}^{i=q}\frac{R}{\prescript{}{i}{d}_{p}}
    \end{bmatrix}$}.
    $}
 \end{equation}
All the derivations up to~\eqref{eq:MIMO_Asm} can be extended to a scenario with different receivers' sizes. The key for such an extension is in~\eqref{eq:SIMO_S}, where each row corresponds to a specific receiver. Thus, all we need to do is to write each line for a receiver with an arbitrary size of the radius. We skip the proof and just write the asymptotic results
 \begin{equation}
 \resizebox{0.88\hsize}{!}{$
     \lim_{t \to \infty} \begin{bmatrix}
        {N}_{1}\left(t \right )\\ 
        {N}_{2}\left(t \right )\\
        \vdots\\
        {N}_{p}\left(t \right )
    \end{bmatrix}
    =
    \resizebox{.67\hsize}{!}{$\begin{bmatrix}
        1 & \frac{R_{1}}{d_{1,2}} & \frac{R_{1}}{d_{1,3}} & \hdots & \frac{R_{1}}{d_{1,p}}\\ 
        \frac{R_{2}}{d_{2,1}} & 1 & \frac{R_{2}}{d_{2,3}} & \hdots & \frac{R_{2}}{d_{2,p}}\\
        \frac{R_{3}}{d_{3,1}} & \frac{R_{3}}{d_{3,2}} & 1 & \hdots & \frac{R_{3}}{d_{3,p}}\\
        \vdots & \vdots & \vdots & \ddots & \vdots\\
        \frac{R_{p}}{d_{p,1}} & \frac{R_{p}}{d_{p,2}} & \frac{R_{p}}{d_{p,3}} &\hdots & 1
    \end{bmatrix}^{\mathrlap{-1}}
    \begin{bmatrix}
        N_T\sum_{i=1}^{i=q}\frac{R_{1}}{\prescript{}{i}{d}_{1}} \\
        N_T\sum_{i=1}^{i=q}\frac{R_{2}}{\prescript{}{i}{d}_{2}} \\
        \vdots \\
        N_T\sum_{i=1}^{i=q}\frac{R_{p}}{\prescript{}{i}{d}_{p}}
    \end{bmatrix}$},
    $}
 \end{equation}
where $R_{i}$ is the radius of the receiver~$\mathcal{R}_{i}$, $i$$\,=\,$$1,\cdots,p$.
\begin{table}[!t]
\begin{center}
\vspace{.1cm}
\caption{Values for system parameters}
\vspace{-.1cm}
\label{tab:param}
\resizebox{0.48\textwidth}{!}{
 \begin{tabular}{|| c | c | c ||}
 \hline
 Variable & Definition & Value \\ [0.5ex] 
 \hline\hline
  $N_{T}$ & Number of released molecules & $10^{4}$ \\ 
  \hline
 $R$ & Receivers radius & $1$ $[\mu$m]\\ 
 \hline
 $d_1$ & Distance between receiver $\mathcal{R}_1$ and transmitter & $6$ $[\mu$m] \\
 \hline
 $D$ & Diffusion coefficient for the signaling molecule & $79.4$ $[\mu\text{m}^2/\text{s}]$ \\
 \hline
\end{tabular}}
\end{center}
\vspace{-.2cm}
\end{table}
\section{Numerical evaluation and results}\label{sec:simulation_results}
This section validates the asymptotic results obtained from the derived asymptotic model with those obtained from the analytical one defined in~\cite{Fardad1}, which was simulated for a long time, for the
SITO and $2$$\,\times\,$$2$ MIMO topologies. With reference to the system of coordinates defined in Fig.~\ref{fig:topology}, in both the two considered scenarios the position of $\mathcal{R}_1$ is fixed in $\left(d_1,0,0\right)$ while that of $\mathcal{R}_2$ varies based on $\Omega$$\,\in\,$$[0,\pi]$ and $d_{C_1C_2}$$\,\in\,$$\{4,8,12\}$. The angle~$\Omega$ is on $xy$\nobreakdash-plane. Table~\ref{tab:param} report the values of the parameters we used to obtain the results shown in the following.
\par Figure~\ref{fig:Sim_result} shows the cumulative number of molecules absorbed by receiver $\mathcal{R}_1$ for the scenario depicted in Fig.~\ref{fig:topology}. The green and red curves are plotted based on analytical model with accurate barycenter position from~\cite{Fardad1} when $t$$\,\in\,$$\{100,300\}\,$s. The blue curve resulted from~\eqref{eq:SITO_ASYM} by assuming that the barycenters are located in the center of the spheres. It can be seen that the asymptotic model with the assumption that the barycenters are located in the center of the FA receivers follows the result obtained by 
the analytical model for a long observation time, which relies on the exact position of the barycenter point.
Figure~\ref{fig:a} shows the cumulative number of absorbed molecules when the distance between the two FA receivers is $4\,\mu\text{m}$. We observe that when $\Omega$$\,=\,$$0$, receiver $\mathcal{R}_2$ is between $\mathcal{R}_1$ and $\mathcal{T}$, thus blocking the ``line-of-sight'' (LOS) between them. In this case the number of absorbed molecules $N_1$ has the lowest value and then it increases as the position of $\mathcal{R}_2$ changes for higher values of $\Omega$. This happens because $\mathcal{R}_2$ absorbs less molecules from the environment, and consequently it has less effect on $\mathcal{R}_1$.
In Fig.~\ref{fig:b} we increased the distance between the receivers. When $\Omega$$\,=\,$$0$°, $\mathcal{R}_2$ is very close to the transmitter and located at $(-2,0,0)$. Receiver $\mathcal{R}_2$ absorbs molecules but in comparison with $\Omega$$\,=\,$$0$° at Fig.~\ref{fig:a}, $N_1$ has higher value because $\mathcal{R}_2$ does not block the LOS. Moreover, for $\Omega$$\,=\,$$90$° and higher the variation of $N_1$ is not so large because $\mathcal{R}_2$ is sufficiently far from both $\mathcal{R}_1$ and transmitter thus, has less effect of $N_1$. 
Finally, in Fig.~\ref{fig:c} the distance is increased to $12\,\mu$m. In this case, we observe that the variation of $N_1$ is not large due to the increase of the distance between $\mathcal{R}_2$ and $\mathcal{R}_1$. When $\Omega$$\,=\,$$0$°, $\mathscr{R}_2$ is closer to the transmitter compared with $\Omega$$\,=\,$$180$° and, therefore, the value of $N_1$ is a little bit lower than that at $\Omega$$\,=\,$$0$°. 
\begin{figure*}[t!]
    \centering
  \subfloat[\label{fig:a_TITO}]{%
       \includegraphics[width=0.33\linewidth]{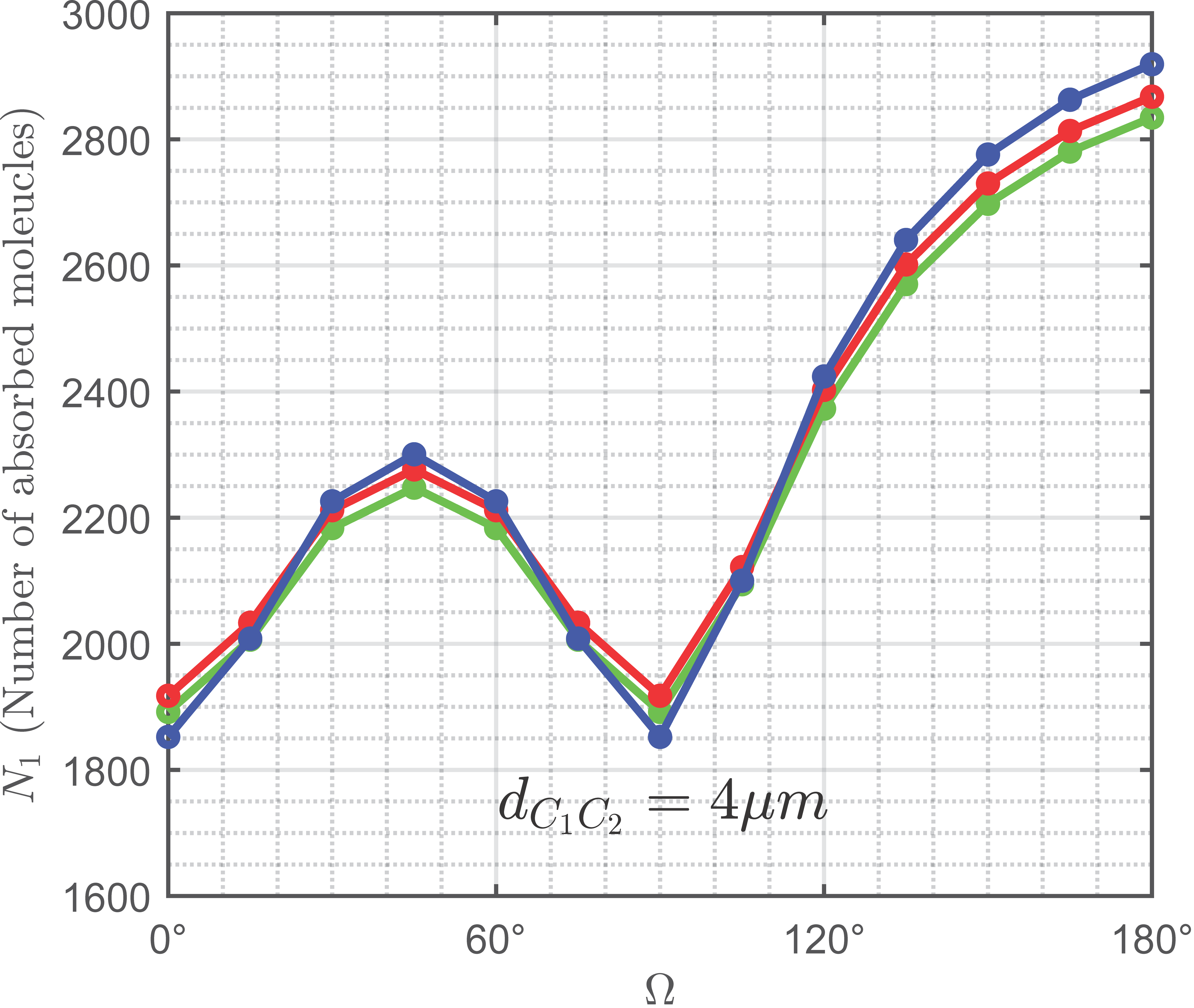}}
  \subfloat[\label{fig:b_TITO}]{%
        \includegraphics[width=0.31\linewidth]{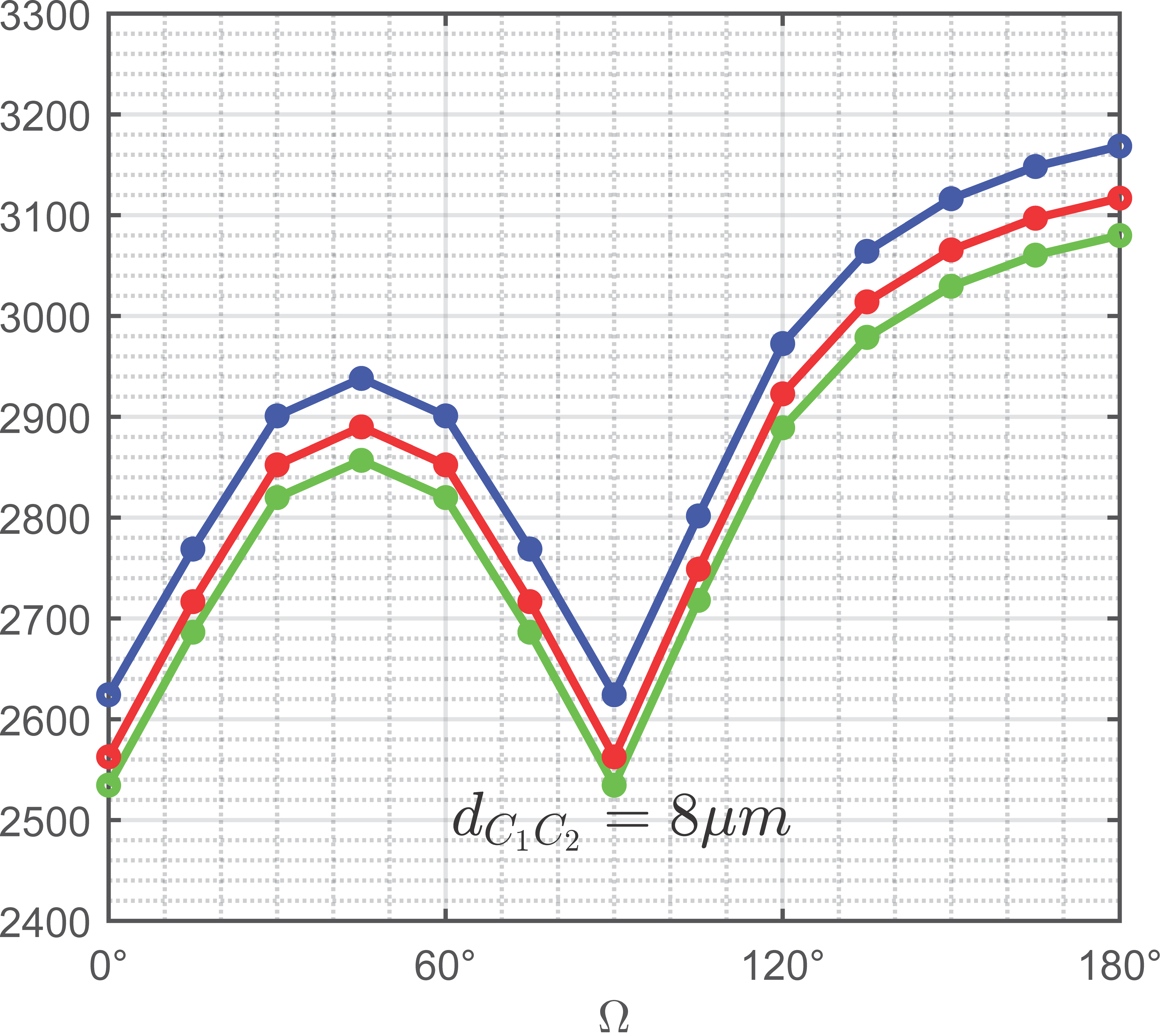}}
  \subfloat[\label{fig:c_TITO}]{%
        \includegraphics[width=0.31\linewidth]{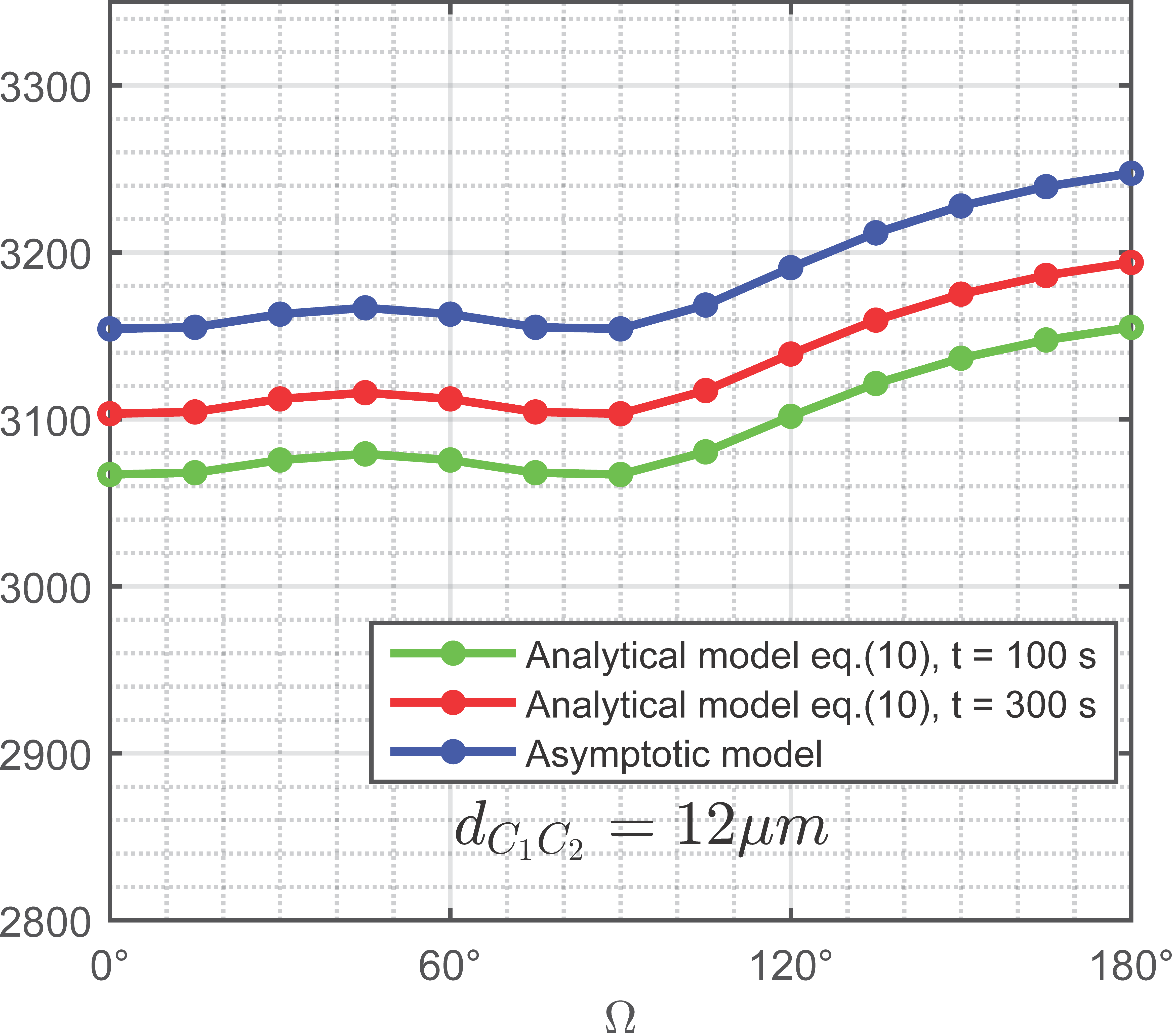}}
  \caption{Cumulative expected number of molecules $N_1(t)$ absorbed by $\mathcal{R}_1$ after $t$$\,\in\,$$\{100,300\}$ seconds, in the scenario of Fig.~\ref{fig:topology_TITO} with $d_1$$\,=\,$$6\,\mu\mathrm{m}$ and $\mathcal{T}_2$$\,=\,$$(6,6,0)\mu\mathrm{m}$, where (a), (b), and (c) corresponds to the same positions of $\mathcal{R}_2$ considered in Fig.~\ref{fig:Sim_result}
. } \label{fig:Sim_result_TIO}
\end{figure*}
\begin{figure}[!t]
    \centering
    \includegraphics[width=0.85\columnwidth]{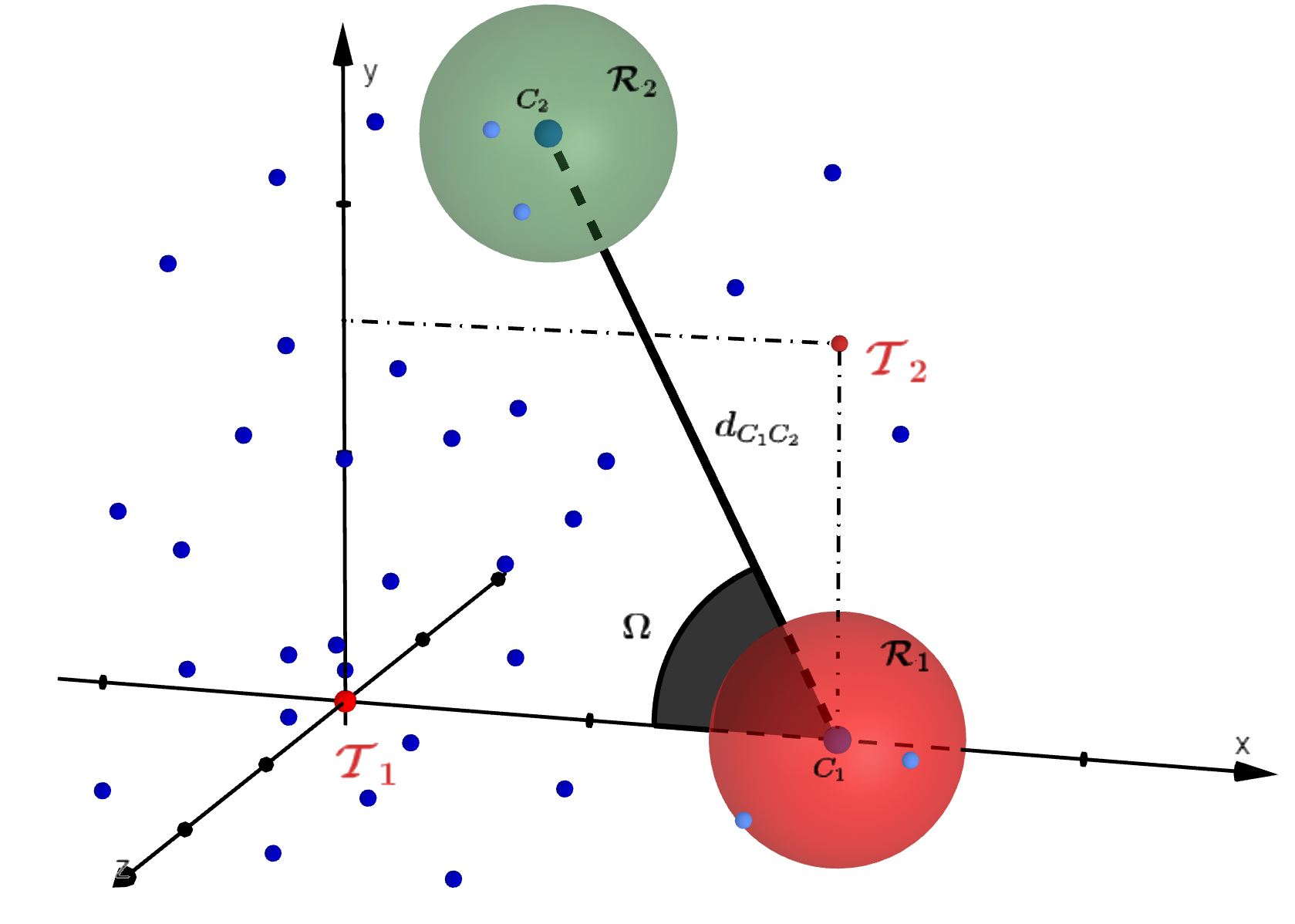}
    \caption{MCvD system with two transmitters and two FA receivers centered at points $C_1$ and $C_2$.}
    \label{fig:topology_TITO}
\end{figure}
\par Figure~\ref{fig:Sim_result_TIO} shows the cumulative number of molecules absorbed by $\mathcal{R}_1$ for the topology depicted in Fig.~\ref{fig:topology_TITO}, where a second pointwise transmitter is present. The results show that our model is able to capture the reciprocal effect of FA receivers in a MIMO scenario. Here, it is assumed that the second transmitter,~$\mathcal{T}_2$ is located at~$(6,6,0)\mu\mathrm{m}$.
The position of the second transmitter was chosen because it can easily highlight the blocking effect associated with the movement of $\mathcal{R}_2$ from $\Omega$$\,=\,$$0$° to $\Omega$$\,=\,$$180$°. As for the previous scenario, in Fig.~\ref{fig:a_TITO} we can observe a good agreement between our asymptotic model and the exact analytical expected cumulative number of absorbed molecules. Figs.~\ref{fig:b_TITO} and~\ref{fig:c_TITO} demonstrate the correctness of the dynamic behavior of the asymptotic model. For these two latter cases we do not have any blocking effect, thus the cumulative number of absorbed molecules is higher and this demonstrates the validity of the asymptotic model.
\section{Conclusion}\label{sec:conclusion}
We derived an analytical model that captures the asymptotic value of the cumulative number of molecules absorbed by each receiver in a diffusive MIMO molecular communication (MC) system. 
The model is validated by comparing it with the results obtained by 
solving the system of equations that describes the exact analytical model that takes into account the reciprocal effect among the fully absorbing (FA) receivers. In the exact analytical model each FA receiver is described as a pointwise negative source of molecules. The best position of the pointwise negative sources is the barycenter defined by the given geometry. We show that in the temporal asymptotic case the effect of the interfering receivers can be described by approximating their barycenters with the centers of the spheres.
\begin{figure}[!t]
    \centering
    \includegraphics[width=0.85\columnwidth]{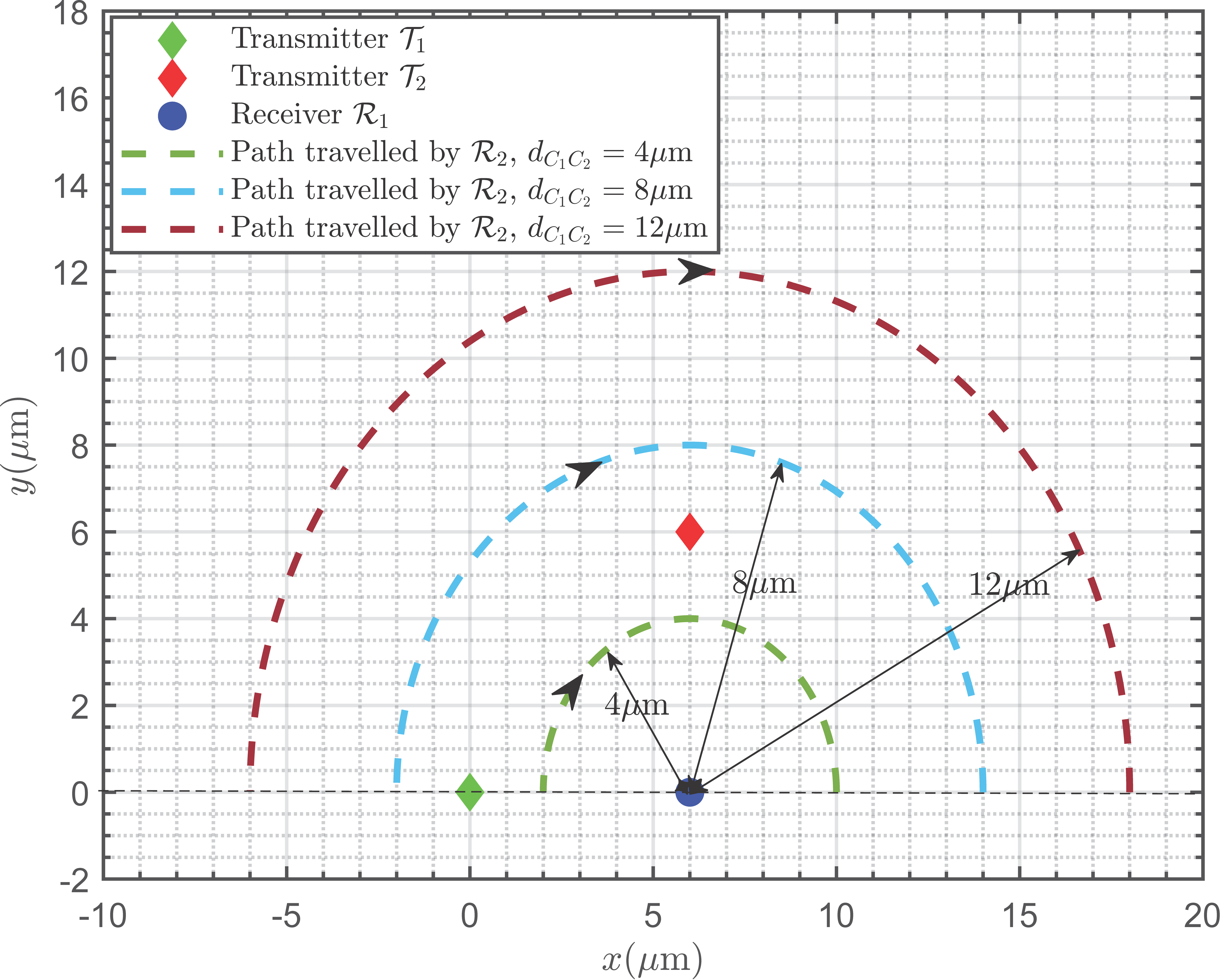}
    \caption{A 2D perspective of the scenario depicted in~\ref{fig:topology_TITO}. The dashed lines represents the travelled path by $\mathcal{R}_2$ for different $d_{C_1C_2}$.}
    \label{fig:xy_TITO}
\end{figure}
\par The proposed asymptotic model opens the ways to the investigation of many problems in diffusive MC systems based on the use of multiple FA receivers such as, target localization and sensing based on the number of absorbed molecules that are measured for a temporal interval that is long enough. 


\ifCLASSOPTIONcaptionsoff
  \newpage
\fi

\bibliographystyle{IEEEtran}
\bibliography{IEEEabrv,pulseshape}

\end{document}